\documentclass[prd,twocolumn,showpacs,preprintnumbers,floats,floatfix,nofootinbib]{revtex4-1}
\usepackage{graphicx}
\usepackage{dcolumn}
\usepackage{latexsym}

\begin{document}

\title{Numerical Evolution of axisymmetric vacuum spacetimes: a code based on the Galerkin method}
\author{H. P. de Oliveira}
\email{oliveira@dft.if.uerj.br}
\author{E. L. Rodrigues}
\email{elrodrigues@uerj.br}
\affiliation{{\it Universidade do Estado do Rio de Janeiro }\\
{\it Instituto de F\'{\i}sica - Departamento de F\'{\i}sica Te\'orica}\\
{\it Cep 20550-013. Rio de Janeiro, RJ, Brazil}}

\date{\today}

\begin{abstract}
We present the first numerical code based on the Galerkin and Collocation methods to integrate the field equations of the Bondi problem. The Galerkin method like all spectral methods provide high accuracy with moderate computational effort. Several numerical tests were performed to verify the issues of convergence, stability and accuracy with promising results. This code opens up several possibilities of applications in more general scenarios for studying the evolution of spacetimes with gravitational waves.
\end{abstract}


\maketitle

\section{Introduction}

In their seminal works Bondi et al \cite{bondi,bondi1} have launched the basis for a detailed description for the dynamics of the exterior axisymmetric spacetime of a bounded source undergoing a process of gravitational wave emission. They have established the suitable form of the metric, the field equations and also exhibited the set of permissible coordinate transformations that preserves the nature of the metric of the Bondi problem. Their most important result was the introduction of the \textit{news function} that determines the rate of mass loss due to gravitational wave extraction, along with providing an invariant characterization of the presence of gravitational radiation. Nonetheless, another important pioneering aspect of this work was the presentation of the field equations in the scheme of characteristics \cite{living_rev_winicour} more than ten years before the formalization of the Cauchy approach \cite{cauchy} of Einstein equations.

The first numerical code to integrate the Bondi equations was developed by Issacson, Welling and Winicour \cite{issacson}, and later improved by a version that avoided instabilities near the vertex \cite{winicour}. Other groups have also presented distinct strategies for constructing their codes as, for instance, using the tetrad formalism \cite{stewart} or combining Cauchy and characteristics evolution that allowed the extension of the Bondi problem to full axisymmetry \cite{southampton}. A detailed description of the numerical schemes for the characteristic evolution can be found in Ref. \cite{living_rev_winicour}, but the common feature shared by all of these codes is that they were constructed using finite difference techniques.

Spectral methods such as Galerkin, Collocation, Tau, etc \cite{spectral} belong to the general class of Weighted Residual Methods \cite{finlayson}. These methods represent an alternative strategy to solve numerically any differential equation for which the solution is approximated by a finite series expansion with respect to a convenient set of analytical functions known as basis functions. This series expansion is such that some quantity - a measure of error or residual - which should be exactly zero is forced to be zero in an approximate sense. To illustrate briefly this important aspect, let us consider a function $f(x)$ that is approximated by a series expansion $f_N(x)=\sum_{k=0}^N\,a_k \psi_k(x)$, where $a_k$ are the unknown modes or coefficients and $\psi_k(x)$ are the basis functions. The modes $a_k$ are determined by demanding that the residual $R_N \equiv f(x)-f_N(x)$ is set to zero in some approximate sense as indicated by,

\begin{equation}
\left<R_N,\phi_j\right> = \int_a^b\,R_N \phi_j w dx =0, \label{eq1.1}
\end{equation}

\noindent where $j=0,1,..,N$, $\phi_j(x)$ are the test functions and $w$ is the weight. The choice of the test function defines the type of spectral method. For instance, if $\phi_k(x)=\psi_k(x)$ with the basis functions satisfying the boundary conditions we have the Galerkin method. On the other hand if $\phi_j(x)=\delta(x-x_j)$, with the $x_j$ being the collocation or grid points, characterizes the Collocation method. In this case the residual vanishes exactly at the collocation points or $R_N(x_j) = 0$.


As a consequence, any spectral method transforms an evolution partial differential equation into a finite set of ordinary differential equations. We highlight two main features of spectral methods: (i) considerable economy of the computational resources to achieve a given accuracy if compared with the finite difference techniques; (ii) the possibility of selecting a coordinate system adapted to the geometry of the problem under consideration that allows the exact treatment of pseudo-singularities present in the chosen coordinates. As a consequence, the use of spectral methods in Numerical Relativity \cite{sm_num_rel} has increased considerably in the last years becoming a viable alternative to the finite difference scheme.

In this work we have developed the first numerical code based on a variant of the Galerkin method known as Galerkin method with numerical integration (G-NI)~\cite{GNI} and the Collocation method~\cite{spectral,boyd,galerkin} to integrate the field equations of the Bondi problem. In Section 2 the field equations are presented along with a brief description of their relevant aspects. The appropriate basis functions for the Galerkin expansions of the metric functions and the description of the numerical implementation are presented in Section 3. Section 4 is devoted to present the numerical tests of the code. Finally, we have made some concluding remarks in Section 5.

\section{The Bondi problem}

The metric for axisymmetric and asymptotically flat spacetimes corresponding to the Bondi problem \cite{bondi1} takes the form

\begin{eqnarray}
& &ds^2=\left(\frac{V}{r}{\rm e}^{2\beta}-U^{2}r^{2}{\rm e}^{2\gamma}\right)du^2 + 2{\rm e}^{2\beta }dudr+ \nonumber \\
& & +2Ur^2 {\rm e}^{2\gamma } dud\theta-r^2 ({\rm e}^{2\gamma }d\theta^2 +{\rm e}^{-2\gamma }{\sin}^{2}\theta d\phi ^{2}), \label{eq1}
\end{eqnarray}

\noindent where $u$ is the retarded time coordinate for which the outgoing null cones are denoted by $u$=constant. The radial coordinate $r$ is chosen such that the surfaces of constant $(u,r)$ have area $4 \pi r^2$, and the angular coordinates $(\theta,\phi)$ are constant along the outgoing null geodesics. The functions $\gamma$, $\beta$, $U$ and $V$ depend on the coordinates $u,r,\theta$ and satisfy the vacuum field equations $R_{\mu\nu}=0$ organized in three hypersurface equations and one evolution equation as shown by Bondi et al \cite{bondi1}. These equations are, respectively, written as,

\begin{widetext}
\begin{eqnarray}
&\beta _{,r}  =  \frac{1}{2}r (\gamma_{,r})^{2} \label{eq2}& \\
\nonumber \\
&\left[r^{4}{\rm e}^{2(\gamma -\beta )}U_{,r}\right]_{,r} = 2r^2\left[r^{2}\left(\frac{\beta}{r^2}\right)_{,r\theta}-\frac{(\sin^2\theta\, \gamma)_{,r\theta}}{\sin^{2}\theta}+2\gamma_{,r}\gamma_{,\theta}\right]& \label{eq3} \\
\nonumber \\
& V_{,r}  =  -\frac{1}{4}r^{4}{\rm e}^{2(\gamma-\beta)}(U_{,r})^{2}+\frac{(r^4\sin\theta U)_{,r \theta}}{2r^2 \sin\theta} + {\rm e}^{2(\beta-\gamma)}\left[1-\frac{(\sin\theta \beta_{,\theta})_{,\theta}}{\sin\theta}+\gamma_{,\theta\theta} \right. & \nonumber\\
\nonumber\\
&+3\cot\theta \, \gamma_{,\theta}-(\beta_{,\theta})^{2} - 2\gamma_{,\theta}(\gamma_{,\theta}-\beta_{,\theta})\Big],&	 \label{eq4} \\
\nonumber \\
& 4r(r\gamma )_{,ur}=\left\{2r\gamma_{,r}V-r^{2}\left[2\gamma_{,\theta}U+\sin\theta\left(\frac{U}{\sin\theta}\right)_{,\theta}\right]\right\}_{,r}-
2r^2\frac{(\gamma_{,r}U\sin\theta)_{,\theta}}{\sin\theta} + \frac{1}{2}r^{4}{\rm e}^{2(\gamma-\beta)}(U_{,r})^{2}&\nonumber \\
&+2{\rm e}^{2(\beta-\gamma)}\left[(\beta_{,\theta})^{2}+\sin\theta \left(\frac{\beta_{,\theta}}{\sin\theta}\right)_{,\theta}\right].& \label{eq5}
\end{eqnarray}
\end{widetext}

\noindent The subscripts $r$, $u$ and $\theta$ denote derivatives with respect to these coordinates. Notice that the evolution equation (\ref{eq5}) is the only one that has derivative with respect to $u$, while the hypersurface equations contain only derivatives in the null hypersurfaces $u=$constant. The regularity conditions of the spacetime at the origin are:

\begin{equation}
\gamma \sim \mathcal{O}(r^2),\;\; \beta \sim \mathcal{O}(r^4),\;\; U \sim \mathcal{O}(r),\;\; V \sim r + \mathcal{O}(r^3). \label{eq6}
\end{equation}

\noindent Also, by imposing smoothness of the symmetry axis, it is necessary that

\begin{equation}
\bar{\gamma} \equiv \frac{\gamma}{\sin^2\theta}, ~ \bar{U} \equiv\frac{U}{\sin\theta} \label{eq7}
\end{equation}

\noindent be continuous at $\theta=0, \pi$. For the sake of convenience we shall consider $\bar{\gamma},\bar{U}$ when formulating the numerical code to solve the field equations.

In the seminal work of Bondi et al \cite{bondi1} the characteristic formulation of the Einstein's equations \cite{living_rev_winicour} was introduced and analyzed for the first time. The evolution scheme is the following: once the initial data $\gamma_0(r,\theta)=\gamma(u_0,r,\theta)$ is specified, the hypersurface equations (\ref{eq2}), (\ref{eq3}) and  (\ref{eq4}) determine the metric functions $\beta$, $U$ and $V$ (modulo integration constants) on the initial null surface $u=u_0$. From these results, the evolution equation (\ref{eq5}) provides $\gamma_{,u}$ on $u=u_0$, and consequently allows the determination of $\gamma$ on the next null surface $u=u_0+\delta u$, and the whole cycle repeats providing the evolution of the spacetime.

According to Ref. \cite{bondi1} (see also \cite{issacson,winicour}) the asymptotic form of the metric functions are,

\begin{eqnarray}
\gamma &=& K(u,\theta) + \frac{c(u,\theta)}{r} + \mathcal{O}(r^{-2}), \label{eq8} \\
\nonumber \\
\beta &=& H(u,\theta) + \mathcal{O}(r^{-2})  \label{eq9}\\
\nonumber \\
U &=& L(u,\theta) + \mathcal{O}(r^{-1})  \label{eq10}\\
\nonumber \\
V &=& \frac{L \sin \theta}{\sin \theta}r^2 + r \mathrm{e}^{2(H-K)}V_1(u,\theta)-2\mathrm{e}^{2H}\,M(u,\theta) + \mathcal{O}(r^{-1}),\nonumber \\ \label{eq11}
\end{eqnarray}

\noindent where $V_{1}$ is related to the functions $H,K,L$, and $M(u,\theta)$ is the Bondi's mass aspect~\cite{bondi1,winicour1}. 
The Bondi mass and the news function are determined by the asymptotic quantities present in Eqs. (\ref{eq8}-\ref{eq11})~\cite{winicour,issacson,winicour1}, and are connected by the Bondi formula:

\begin{equation}
\frac{d M_B(u)}{d u} = -\frac{1}{2}\,\int_{0}^\pi\,\frac{\mathrm{e}^{2 H}}{\omega} N^2 \sin \theta d \theta,\label{eq11_1}
\end{equation}

\noindent where $M_B(u)$ is the Bondi mass, $N(u,\theta)$ is the news function and $\omega$ is a function that depends of the gauge we are considering~\cite{issacson}. The corresponding expressions for these quantities in terms of the asymptotic quantities are presented in Section 4.

\section{The Galerkin-Collocation approach}

As a type of spectral method, the G-NI method \cite{galerkin} establishes that solutions of differential equations are approximated by finite series expansion with respect to a basis functions, in which each basis function satisfies the boundary conditions. In order to implement the Galerkin method to integrate the field equations (\ref{eq2} - \ref{eq5}), we need to establish appropriate series expansions for the metric functions $\gamma$, $\beta$, $U$ and $V$.

The proposed Galerkin expansion for $\bar{\gamma}$ has the form,

\begin{eqnarray}
\bar{\gamma}_a(u,r,x) = \sum_{j=0}^{N_x}\sum_{k=0}^{N_r}\,a_{kj}(u)\Psi^{(\gamma)}_k(r)P_j(x), \label{eq12}
\end{eqnarray}

\noindent where the subscript $a$ indicates an approximation of the actual function $\bar{\gamma}$, and the angular coordinate was changed according to $x=\cos\theta$. $N_r$ and $N_x$ are the truncation orders of the radial and angular expansions, respectively; $a_{kj}(u)$ are the modes that depend on the retarded time $u$. The Legendre polynomials, $P_j(x)$, are the natural choice for the angular basis that are regular at $\theta=0,\pi$ or $x = \pm 1$. The radial basis functions, $\Psi^{(\gamma)}_k(r)$, are expressed by a suitable linear combination of the rational Chebyshev polynomials \cite{boyd} (see Appendix A) in order to satisfy the boundary conditions (\ref{eq6}) and (\ref{eq8}), yielding

\begin{eqnarray}
\begin{array}{l l}
\Psi^{(\gamma)}_k(r)  \sim \mathcal{O}(r^2)\;\; \mathrm{near}\,r = 0 \label{eq13} \\
\\
\Psi^{(\gamma)}_k(r) \sim  \mathrm{constant}\; +\; \mathcal{O}(r^{-1}), \;\; \mathrm{near}\; \mathfrak{J^+},\\
\end{array} 
\end{eqnarray}

\noindent valid for all $k=0,..,N_r$. The asymptotic expression of $\gamma$ given by Eq. (\ref{eq8}) is recovered by the Galerkin expansion (\ref{eq12}) in which the functions $K(u,x)$ and $c(u,x)$ are expressed in terms of the modes $a_{kj}$ and the Legendre polynomials.

The Galerkin expansions for the metric functions $\bar{U}$ is given by,

\begin{eqnarray}
\bar{U}_a(u,r,x) =\sum_{j=0}^{M_x}\sum_{k=0}^{M_r}\,b_{kj}(u)\Psi^{(U)}_k(r)P_j(x) 
\label{eq14}
\end{eqnarray}

\noindent where $b_{kj}(u)$ are the modes, and $\Psi_k^{(U)}(r)$ (see Appendix A) represents the radial basis functions that obey the boundary conditions (\ref{eq6}) and (\ref{eq10}). We have followed Ref. \cite{thesis} and introduced the function $S(u,r,x)$ such that $V \equiv r + r^2 S$. As a consequence, $S$ obeys the same boundary conditions of $\bar{U}$, which lead us to propose the following Galerkin expansion,

\begin{eqnarray}
S_a(u,r,x) = \sum_{j=0}^{M_x}\sum_{k=0}^{M_r}\,s_{kj}(u)\Psi^{(U)}_k(r)P_j(x), \label{eq15}
\end{eqnarray}

\noindent where $s_{kj}$ are the modes associated to $S$. The expansion for the function $\beta \equiv (1-x^2)^2 \bar{\beta}$ follows the same scheme as above,

\begin{eqnarray}
\bar{\beta}_a(u,r,x) = \sum_{j=0}^{M_x}\sum_{k=0}^{M_r}\,c_{kj}(u)\Psi^{(\beta)}_k(r)P_j(x). \label{eq16}
\end{eqnarray}

\noindent where $c_{kj}(u)$ are the unknown modes, and the radial basis functions $\Psi_k^{(\beta)}(r)$ (see Appendix A) satisfy the boundary conditions given by Eqs. (\ref{eq6}) and (\ref{eq9}).

We proceed by substituting the approximate functions $\bar{\beta}_a$, $\bar{\gamma}_a$, $\bar{U}_a$ and $S_a$ into the field equations to obtain the corresponding residual equations. For the sake of illustration, let us consider the residual equation associated to the hypersurface Eq. (\ref{eq2}),

\begin{eqnarray}
\mathrm{Res}_{\bar{\beta}}(u,x,y) = \bar{\beta}_{a,y} - \frac{1}{4}(1-y^2) \bar{\gamma}_{a,y}^2 \label{eq17}
\end{eqnarray}

\noindent In this expression we introduced the compactified radial variable $y=(r-1)/(r+1)$ in which $-1 \leq y \leq 1$. Since the expansions for $\bar{\beta}$ and $\bar{\gamma}$ are approximations, the residual equation does not vanish exactly, but it is expected that the residual equation approaches to zero as the truncation orders are increased.

According to the Method of Weighted Residuals~\cite{finlayson} from which the Galerkin method is one of its variants, the unknown modes are chosen such that the residual equation is forced to be zero in an average sense \cite{finlayson}. This means that the inner products of the residual equation are set to zero, or equivalently,

\begin{equation} \left<\mathrm{Res}_{\bar{\beta}},\bar{\psi}_{kj}\right>=\int_{\mathcal{D}}\,\mathrm{Res}_{\bar{\beta}}\, \bar{\psi}_{kj}(\mathbf{r}) w d^2\mathbf{r} = 0, \label{eq18}
\end{equation}

\noindent where $\mathrm{Res}_{\bar{\beta}}$ is the residual equation, $\bar{\psi}_{kj}(\mathbf{r})$ are the test functions and $w$ the weight function. We have chosen to use the Collocation method that implies in setting $\bar{\psi}_{kj}(\mathbf{r}) = \delta(y-y_k)\,\delta(x-x_j)$ and $w=1$, where $y_k$ and $x_j$ are the collocation or the grid points given by:

\begin{eqnarray}
\begin{array}{l l}
x_k =-1,\,\mathrm{zeros\,of}\, \frac{d P_{N_x}}{d x}, 1,\\
\\
y_k=\cos\left(\frac{k \pi}{M_r+1}\right), k=0,1,..,M_r+1,
\label{eq19}
\end{array}
\end{eqnarray}

\noindent which are the Legendre-Gauss-Lobatto and Chebyshev-Gauss-Lobatto points \cite{spectral,boyd}, respectively. Therefore, the vanishing of the inner products implies that the residual equation is forced to be zero exactly at the collocation points. In the specific case of the residual equation (\ref{eq17}), we have

\begin{eqnarray}
\mathrm{Res}_{\bar{\beta}}(u,x_j,y_k) = 0,
\label{eq20}
\end{eqnarray}

\noindent with $k=0,1,..,M_r$ and $j=0,1,..,M_x$.

In principle, these relations constitute a set of $(M_x+1) \times (M_r+1)$ algebraic equations connecting the modes $c_{lm}$ with $a_{lm}$. However, large expressions for $\bar{\beta}_a$ and $\bar{\gamma}_a$ resulting from higher truncation orders produce large algebraic relations if expressed in terms of the unknown modes $a_{kj},c_{kj}$. Thus, the numerical manipulation of the algebraic system would be computationally very costly. To overcome this difficulty we express the set of equations (\ref{eq20}) as,

\begin{eqnarray}
(\bar{\beta}_{a,y})_{kj}=(\bar{\beta}_{,y})_{kj} = \frac{1}{4}\,(1-y_k^2)\,(\bar{\gamma}_{,y})^2_{kj},\label{eq21}
\end{eqnarray}

\noindent where $(\bar{\beta}_{,y})_{kj}$ and $(\bar{\gamma}_{,y})_{kj}$ are the grid values of $\frac{\partial \bar{\beta}}{\partial y}$ and $\frac{\partial \bar{\gamma}}{\partial y}$, respectively. The determination of the grid values $(\bar{\beta}_{a,y})_{kj}$ implies in the determination of the modes $c_{lm}$ since both quantities are linearly related through,

\begin{eqnarray}
(\bar{\beta}_{,y})_{kj} = \sum_{l=0}^{M_r}\,\sum_{m=0}^{M_x}\,c_{lm}(u) \left(\frac{\partial \Psi^{(\beta)}_l}{\partial y}\right)_{y=y_k} P_m(x_j).\nonumber\\
\label{eq22}
\end{eqnarray}

\noindent \noindent for all $k=0,1,..M_r,j=0,1,..,M_x$. Therefore, once the modes $c_{lm}$ are known the function $\bar{\beta}_a$ is reconstructed.

\begin{figure}[ht]
\rotatebox{0}{\includegraphics*[scale=0.35]{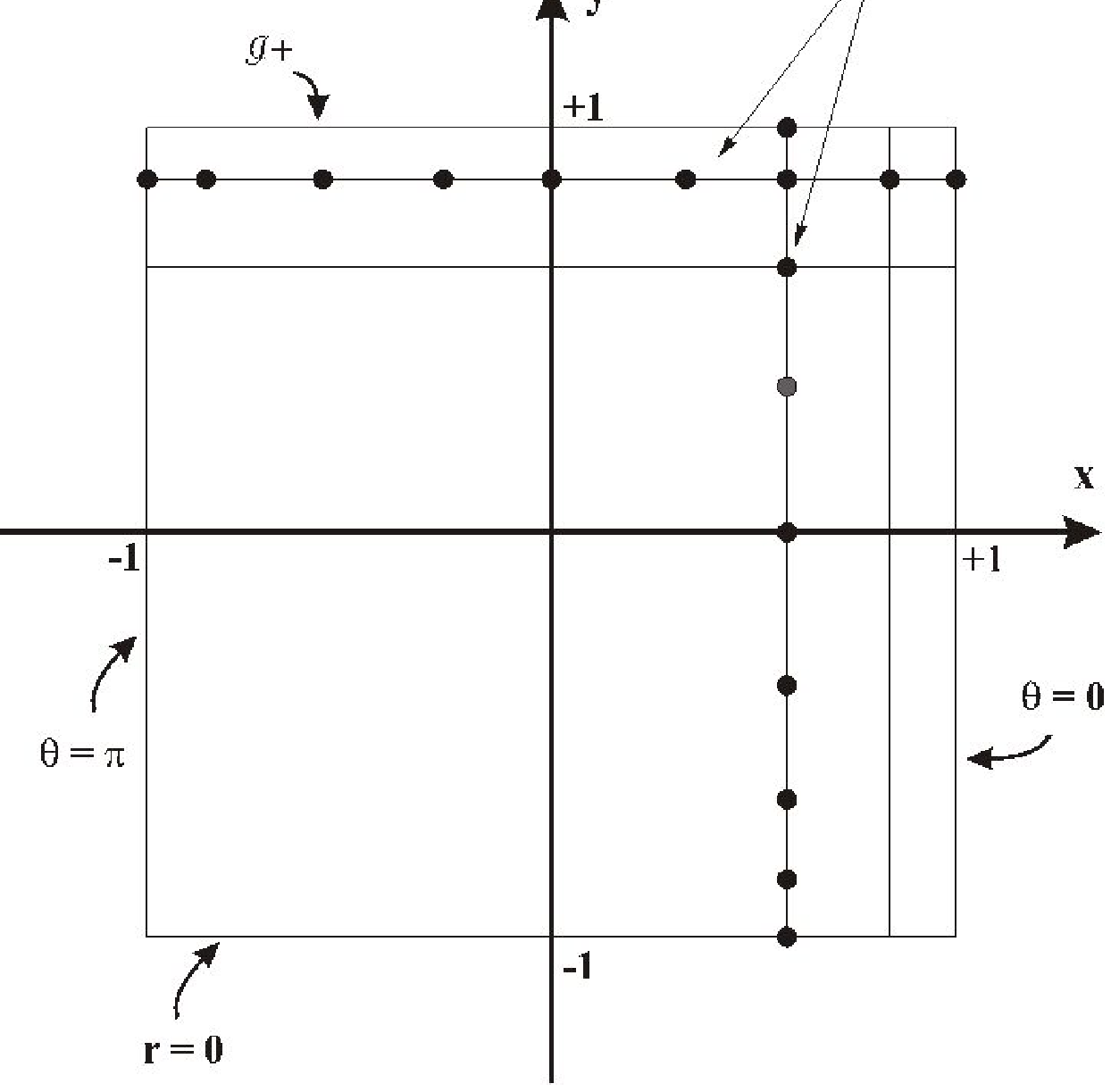}}
\vspace{-2cm}
\caption{Illustration of the collocation or grid points in the spatial domain. The line $y=-1$ represents the origin $r=0$ and the future null infinity the line $y=1$. Notice that the grid points are not regularly spaced.}
\end{figure}

A similar strategy was adopted when taking into account the hypersurface equation (\ref{eq3}). However, instead of using only the representation of the functions given by their grid values, we have also considered the spectral representation that is based on the corresponding unknown modes. In this case we have obtained a set of algebraic equations connecting the grid values $(\bar{U}_{,y})_{kj}$ with the modes $a_{lm}$, the grid values $(\bar{\gamma})_{kj}$, $(\bar{\beta})_{kj}$ and the grid values of their derivatives with respect to $x$ and $y$. Then, the approximate function $\bar{U}_a$ is determined at each slice $u=\mathrm{constant}$ once the grid values $(\bar{U}_{,y})_{kj}$ are known. With respect to the last hypersurface equation (\ref{eq4}), the result of applying a similar treatment is a set of algebraic equations for the grid values $S_{kj}$. In Fig. 1 we show schematically the spatial domain in terms of the coordinates $x$ and $y$ together with the $(M_r+1) \times (M_x+1)$ grid points.

The last step is to consider the evolution equation (\ref{eq5}). As we are going to see in the next Section it is not possible to set $N_x=M_x, N_r=M_r$, so that we can not treat Eq. (\ref{eq5}) using collocation method as done for the hypersurface equations. We propose the use of the G-NI method to establish that the residual equation must vanish in an average sense as indicated by Eq. (\ref{eq18}). Two steps were adopted: first the test functions are no longer delta Dirac functions, but the same as the basis functions, $\Psi_k^{(\gamma)}(r) P_j(x)$; second, the inner products are evaluated using the quadrature formulas. In this case we have,

\begin{eqnarray}
\left<\mathrm{Res_{\gamma}},\Psi_{k}^{(\gamma)}(r) P_{j}(x)\right> &&\approx
\sum_{l=0}^{M_r}\,\sum_{m=0}^{M_x}\,(\mathrm{Res}_{\gamma})_{lm} \Psi_{k}^{(\gamma)}(y_{l}) \times \nonumber \\
&&P_{j}(x_{m})v_m w_l = 0,\label{eq23}
\end{eqnarray}

\noindent where $k=0,1..,N_r,\,j=0,1,..,N_x$; $(\mathrm{Res}_{\gamma})_{lm}$ denotes the values of the residual equation at the grid points (\ref{eq19}), and $v_m,w_l$ are the discretized weights~\cite{spectral,boyd}. These equations yield a system of $(N_r+1)\times (N_x+1)$ ordinary differential equations for the modes $a_{kj}(u)$ schematically represented by,

\begin{eqnarray}
\frac{d{a}_{kj}}{du} = F_{kj},  \label{eq24}
\end{eqnarray}

\noindent where each $F_{kj}$ is a nonlinear function of the modes $a_{lm},b_{lm},s_{lm}$, the grid values of $\bar{\gamma},\bar{U},\bar{\beta},S$ and the grid values of their first derivatives with respect to $x$ and $y$.
Therefore, the field equations are reduced to a system of ordinary differential equations for the modes $a_{kj}(u)$ and three sets of algebraic equations for the grid values $\bar{\beta}_{kj}$, $(\bar{U}_{,y})_{kj}$ and $S_{kj}$. These sets of algebraic equations together with the dynamical system (\ref{eq24}) were generated in Maple that allowed an efficient symbolic manipulation of the field equations. The evolution scheme starts with the initial data $\gamma_0(r,x)=\gamma(u_0,r,x)$ from which the initial modes $a_{lm}(u_0)$ are fixed. Solving the algebraic equations allows the determination of the approximate functions $\bar{\beta_a}$, $\bar{U_a}$ and $S_a$ at $u=u_0$. These functions provide the grid values and modes necessary to obtain the modes $a_{lm}$ in the next slice $u+\delta u$ using the dynamical system $(\ref{eq24})$. The process repeats generating, in this way, the numerical solution of the system. 

\section{Code tests}

We now present the numerical tests \cite{winicour,thesis,gomez} to check the accuracy and convergence of our algorithm. We have implemented second-order Runge-Kutta integrators in Python and Maple with fixed and variable stepsize to evolve the dynamical system (\ref{eq24}). The integrator solves the algebraic equations at each step and evaluates the values and modes necessary to evolve forward the system. All numerical experiments were performed in a desktop with Intel \textregistered Core \texttrademark i7 processor (3.33 GHz) with 24 MB of RAM memory.

The first test consists in evolving small amplitude gravitational waves and compare it with the exact evolution determined by the linearized field equations. In the second test we have considered the evolution of the interior SIMPLE solution~\cite{winicour} matched with an asymptotically flat patch. The third test is a verification of the global energy conservation according to the Bondi formula.

\subsection{Linearized waves.}

The linear regime describes the dynamics of very weak disturbances of the spacetime and is characterized by $\beta \simeq 0$ and $V \simeq r$. In this way the field equations are reduced to the equations for the metric functions $\gamma$ and $U$, where according to Refs.~\cite{bondi1,winicour}, these equations are equivalent to a flat scalar wave equation, $\Box \Phi = 0$. Since $\Phi(u,r,x)$ is related to $\gamma$ and $U$~\cite{winicour}, once the exact solution of the wave equation is known the corresponding exact expressions for $\gamma$ and $U$ are also known.

Let us consider the exact linearized solution corresponding to the harmonic mode $l=6$ given by (see Appendix B),

\begin{eqnarray}
\bar{\gamma}_{[6]} =\frac {5A_0 (693x^4-378x
^2+21)r^6 (2u+r+2)}{12(u+1)^8(u+2r+1)^7},\label{eq25}
\end{eqnarray}

\noindent where $A_0$ is an arbitrary constant. In order reproduce this exact solution with the numerical code we have considered $\gamma_0(r,x) = \gamma_{[6]}(u=0,r,x)$ as the initial data and set $A_0=1.0 \times 10^{-5}$. The error between the exact and numerical solutions is expressed by the $L_2$-norms of the gravitational wave shear $\bar{K}(u,x)=K/(1-x^2)$ (cf. Eq. (\ref{eq8})) and the asymptotic function $\bar{L}=L/\sqrt{1-x^2}$ (cf. Eq. (\ref{eq10})). These norms were evaluated for several truncation orders $N_r$ and their maximum values plotted in Fig. 2. In both cases we notice the exponential decay of the $L_2$-norms together with the saturation due to the round off error occurring for $N_r \geq 16$ and $N_r \geq 18$, respectively.

\begin{figure}[htb]
\rotatebox{0}{\includegraphics*[scale=0.30]{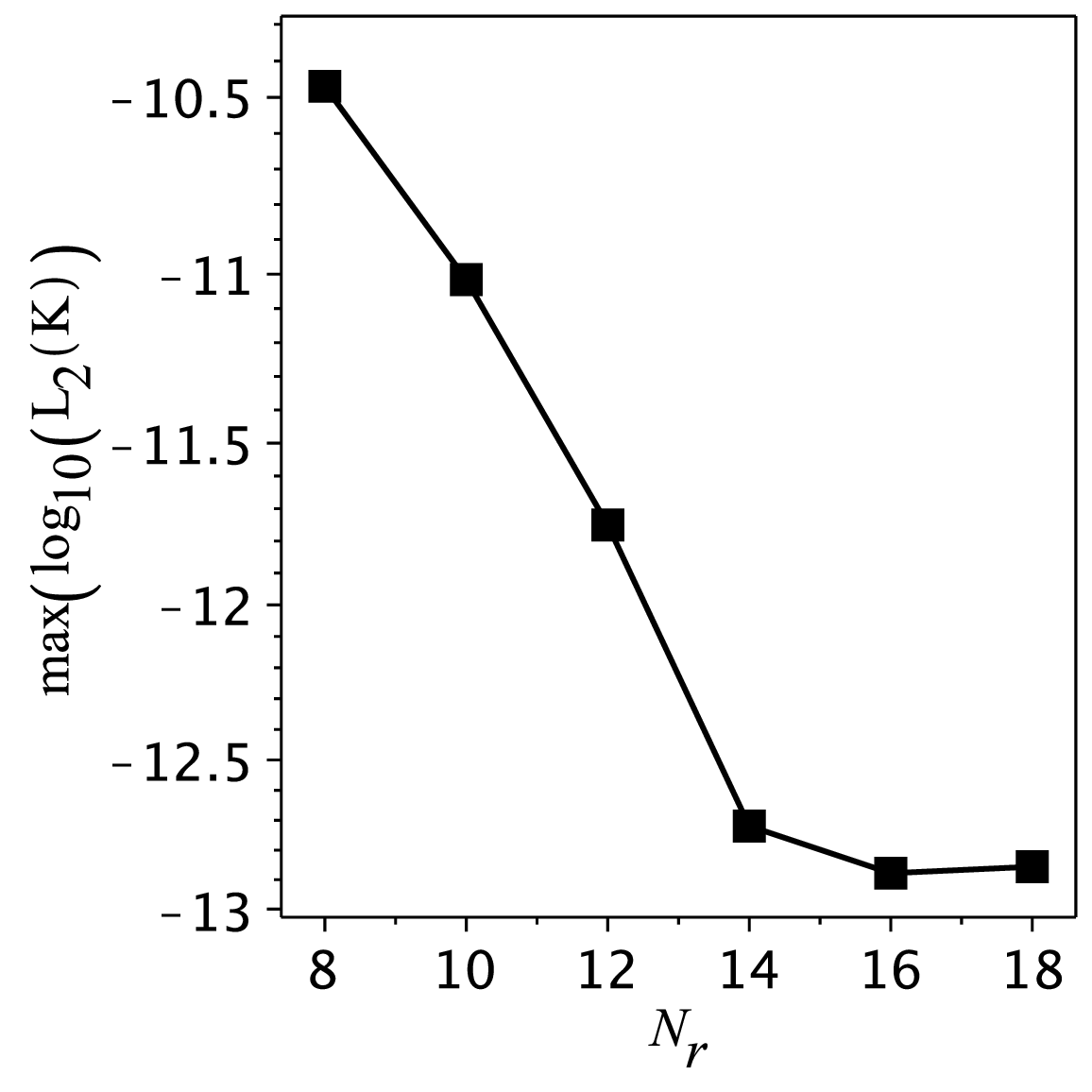}}
\rotatebox{0}{\includegraphics*[scale=0.30]{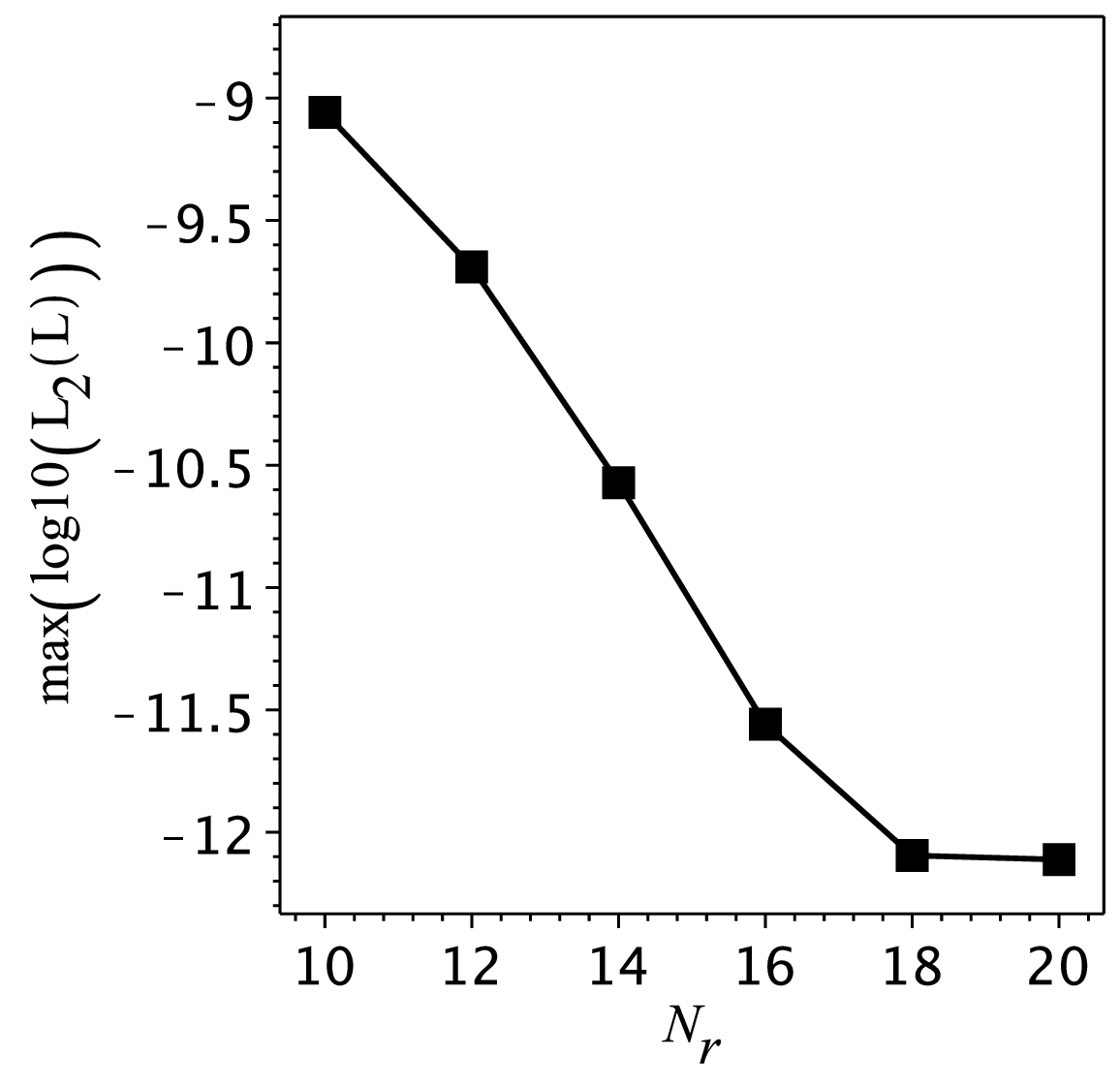}}
\caption{Log-linear plots of the maximum $L_2$ norms between the exact and approximate $\bar{K}$ and $\bar{L}$ associated to the mode $l=6$ (see the Appendix B) evaluated at several truncation orders. Both measures of error decay exponentially until the saturation to the round off error is achieved. In the present numerical experiments we have set $M_r=N_r+2,M_x=N_x+2$, $N_x=4$ and $A_0=10^{-5}$.}
\end{figure}

In the numerical experiments exploring the linear regime, some comments about the choice of the truncation orders $N_x,N_r,M_x,M_r$ are pertinent. We  have considered  $(N_r,N_x)$ (see Eq. \ref{eq12}) as the basic or seed truncation orders from which $(M_r,M_x)$ are related. Due to the common choice of the angular basis function and the forms of the radial basis functions, the linearized equations dictate that $M_r \geq N_r$ and $M_x \geq N_x+1$. Both restrictions provide the minimum values for $M_r,M_x$ that guarantee the stability of the code. In a certain sense, these conditions might be viewed as similar to the Courant-Friedrichs-Lewis (CFL) condition that dictates a suitable time step to assure the stability of the code when the finite difference technique is applied.  For the numerical experiments related to the linear domain we have fixed $N_x=4$ and set $M_r=N_r+2,M_x=N_x+2$, but other choices can be done for further numerical experiments.

\subsection{The SIMPLE solution.}
\begin{figure}[htb]
\rotatebox{0}{\includegraphics*[height=5.7cm,width=6.7cm]{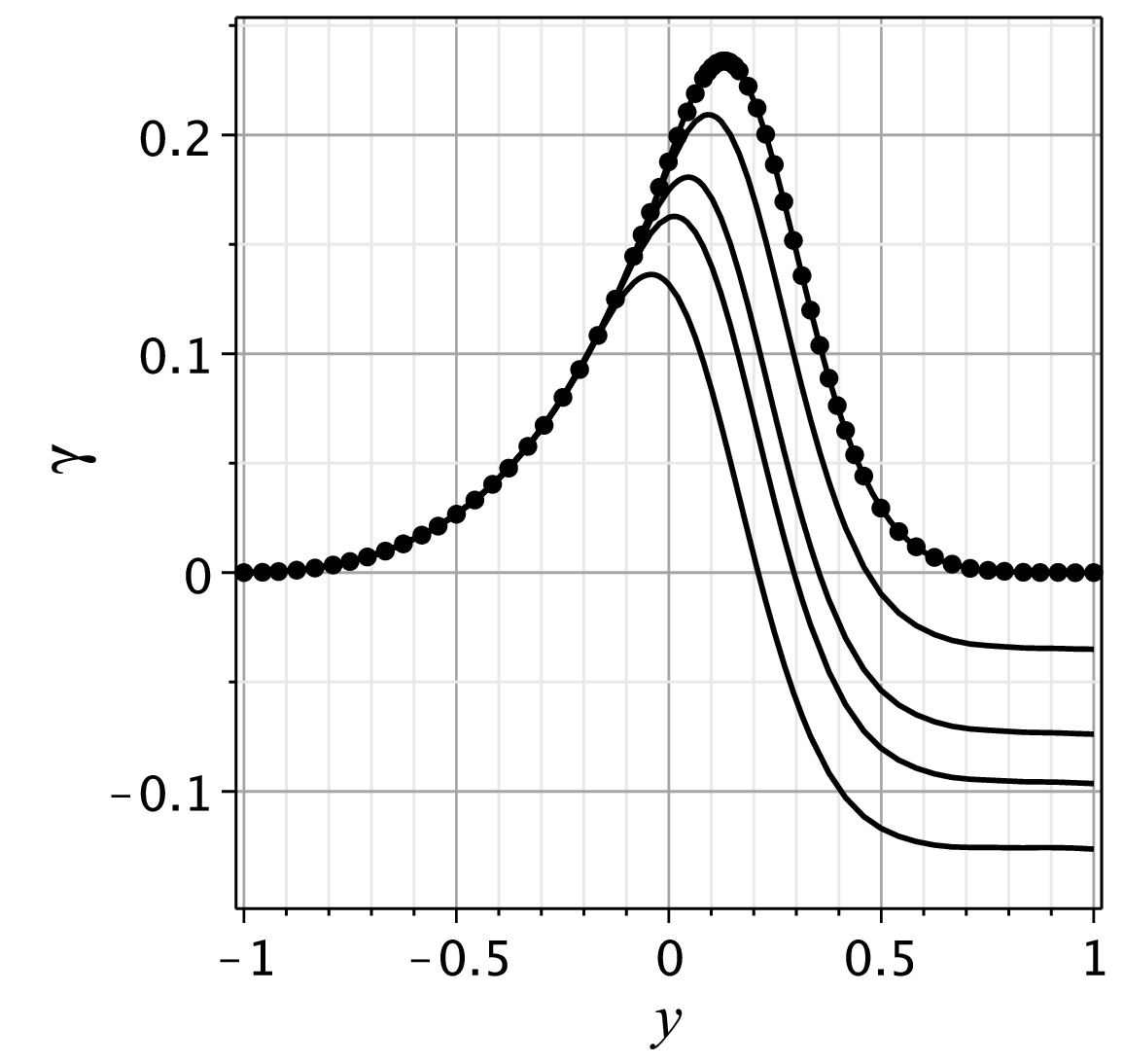}}
\rotatebox{0}{\includegraphics*[height=5.7cm,width=6.7cm]{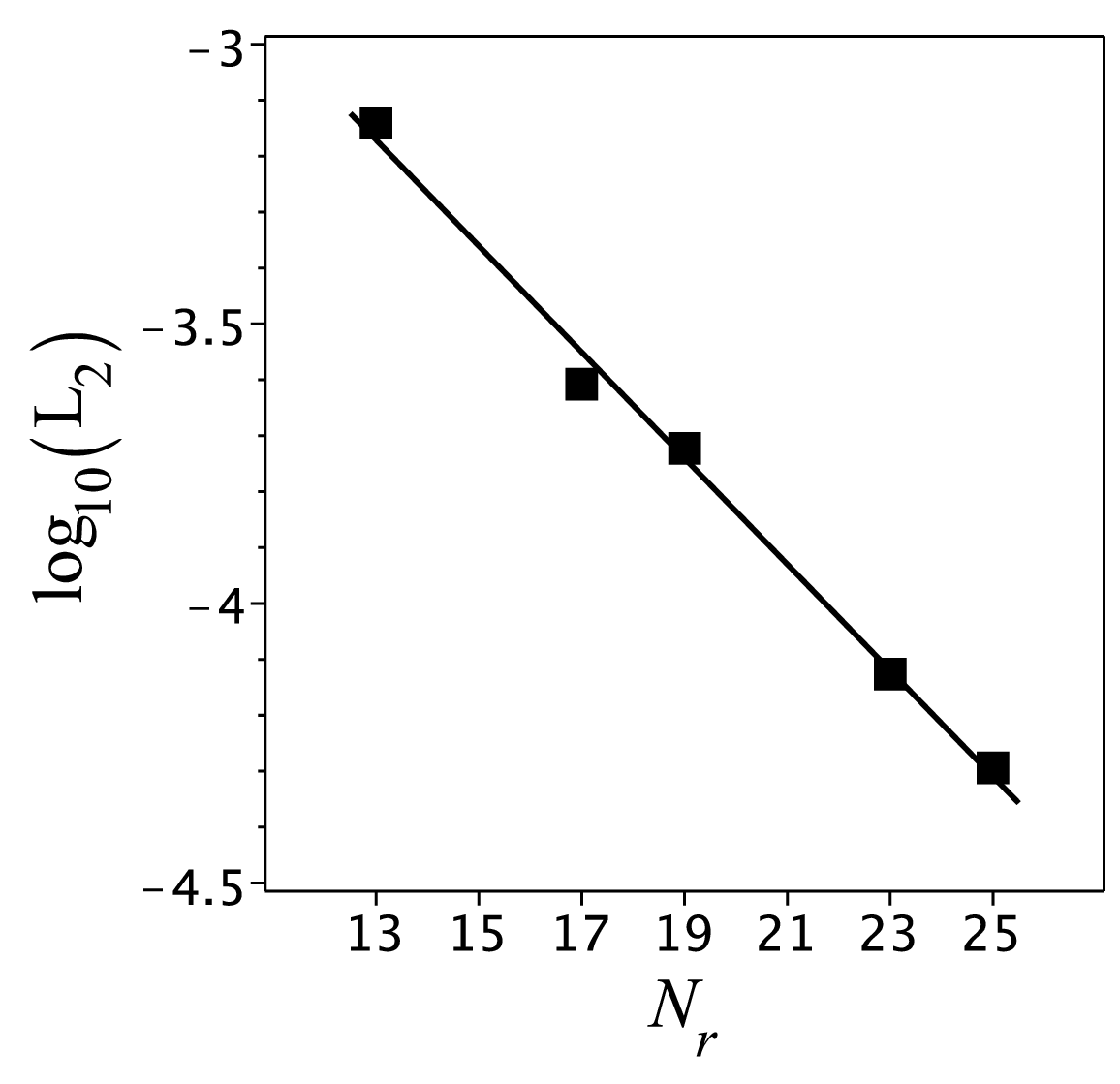}} \vspace{-0.2cm}
\caption{(a) Projections of $\gamma$ on the plane $x=0$ evaluated at several instants. At $u=0$ the dotted curve corresponds to the initial data (\ref{eq27}) and the continuous curves from up to down were obtained numerically. The matching of the asymptotically flat portion of the data to the SIMPLE solution with $A_0=1.0$  is done at $y_m=0$ or $r=1$. The last curve is evaluated at $u=0.4$. (b) Exponential decay of the $L_2$-norm (\ref{eq28}) evaluated at $u=0.3$.}
\end{figure}

The second code test explores the evolution in the nonlinear regime using the SIMPLE solution~\cite{winicour} which is the only known exact static solution of the Bondi equations given by,

\begin{eqnarray}
&&\mathrm{e}^\gamma = \frac{1}{2}(1+\Sigma),\;\mathrm{e}^{2\beta} = \frac{(1+\Sigma)^2}{4 \Sigma},\; U = -\frac{a^2 r \sin \theta \cos \theta}{\Sigma},\nonumber \\
\nonumber \\
&& V = \frac{r}{\Sigma}(2 a^2 r^2 \sin^2 \theta - a^2 r^2 + 1),\label{eq26}
\end{eqnarray}

\noindent where $\Sigma = \sqrt{1+a^2 r^2 \sin^2 \theta}$ and $a$ is a free scale parameter. This solution is not asymptotically flat and therefore cannot be used as initial data. We have followed the procedure outlined in Ref.~\cite{winicour} that consists in constructing an initial data using the SIMPLE solution restricted to a region $r \leq r_c$ by smoothly pasting asymptotically flat null data to it at $r=r_c$. Hence, we proposed the following initial data,

\begin{eqnarray}
(1-x^2) \bar{\gamma}_0(r,x) = \left\{
\begin{array}{l l}
\ln\left[\frac{1}{2}(1+\Sigma)\right] \;\; 0 \leq r \leq r_c \\
\\
  \frac{1}{\left[1+(r-r_c)^2\right]^2}\,\ln\left[\frac{1}{2}(1+\Sigma)\right], \;\; r \geq r_c, \label{eq27}\\
\end{array} \right.
\end{eqnarray}

\noindent where we selected $r_c=1$. The test consists in verifying how accurate the code is able to reproduce the interior static solution inside a region not affected by the nonstatic exterior portion of the data, since the matching boundary propagates along an ingoing null hypersurface. Fig. 3(a) shows the initial data together with the numerical solution and some profiles evaluated at distinct instants. To quantify the error we have used the $L_2$-norm of $\gamma$,

\begin{eqnarray}
L_2(\gamma) = \left(\frac{1}{2}\,\int_{-1}^1\,\int_{0}^{r_s}\,(\gamma_{\mathrm{exact}}-\gamma_a)^2 dr\,dx\right)^{1/2} \label{eq28}
\end{eqnarray}

\noindent where $r_s=2/3$ delimits the region preserved by the SIMPLE solution denoted by $\gamma_{\mathrm{exact}}$. We have evolved the initial data (\ref{eq27}) using the following truncation orders: $N_r=13,17,21,25$ with fixed $N_x=6$, and in each case $M_r=N_r+2$, $M_x=N_x+2=8$. It is clear from Fig. 3(a) that the interior static solution is preserved to a graphical accuracy. Fig. 3(b) shows the exponential decay of the error evaluated at $u=0.3$ when the truncation order is increased, which is an strong indication of the rapid convergence of the algorithm.


\subsection{Global energy conservation. The Bondi mass.}

The last test is to check the global energy conservation obtained after integrating the Bondi formula (\ref{eq11_1}),

\begin{equation}
C(u)=M_B(u) - M_B(u_0) + \frac{1}{2}\,\int_{-1}^1 dx\,\int_{u_0}^u \frac{\mathrm{e}^{2H}}{\omega} N^2 du = 0. \label{eq29}
\end{equation}

\noindent Here $M_B(u)$ is the Bondi mass, $H(u,x)$ is given by Eq. (\ref{eq9}) and $N(u,x)$ is the news function. The conformal factor $\omega$ depends on the gauge we are adopting and arises by connecting the metric of the two geometry of a unit sphere in the standard Bondi coordinates with the similar expression in the present coordinate system, or $d\hat{s}^2_B=d\theta_B^2+\sin^2\theta_Bd\phi_B^2=\omega^2\,(\mathrm{e}^{2K}d\theta^2+ \sin^2\theta\mathrm{e}^{-2K}d\phi^2)$, which yields,

\begin{equation}
\omega=\frac{2 \mathrm{e}^{K}}{(1+x)\mathrm{e}^\Delta+(1-x)\mathrm{e}^{-\Delta}}, \label{eq30}
\end{equation}

\noindent where

\begin{equation}
\Delta= \int_0^x\frac{\mathrm{e}^{2K}-1}{1-x^2}dx.  \label{eq31}
\end{equation}

\noindent It is necessary to determine the functions $\omega(u,x)$ and $\Delta(u)$ for the Bondi formula. For this we have reexpressed $K(u,x)$ as,

\begin{equation}
\mathrm{e}^{2 K}-1 = (1-x^2)\,\sum_{j=0}^{\bar{M}}\,\alpha_j T_j(x),\label{eq32}
\end{equation}

\noindent where $\bar{M} > N$, $T_j(x)$ is the Chebyshev polynomial of jth order, and $\alpha_j,j=0,1,..,\bar{M}$ are the modes. These modes are related to $a_{kj}$ according with the expression for $K(u,x)$ obtained from (\ref{eq12}), that is,

\begin{eqnarray}
K(u,x)=(1-x^2)\,\lim_{r \rightarrow \infty}\,\sum_{k,j=0}^{N_r,N_x} a_{kj}(u)P_j(x)\Psi^{(\gamma)}_k(r). \nonumber
\end{eqnarray}

\noindent From Eq. (\ref{eq32}), the integral (\ref{eq31}) can be done analytically providing the function $\Delta(u)$ and consequently the conformal factor $\omega$.

The news function $N(u,x)$ can be written as~\cite{winicour}

\begin{eqnarray}
N&=&\mathrm{e}^{-2H} c_{,u} - \mathrm{e}^{-2H} \frac{(\sqrt{1-x^2} c^2 L)^\prime}{2c}
 + \frac{1}{2}\mathrm{e}^{-2(K+H)} \times \nonumber \\
 & & \omega (1-x^2)\left[\frac{(\omega\mathrm{e}^{2H})^\prime}{\omega^2}\right]^\prime. \label{eq33}
\end{eqnarray}

\noindent The Bondi mass depends directly on the mass aspect $M(u,x)$ (cf. Eq. (\ref{eq11})) and the terms arising from the gauge under consideration. Following Ref.~\cite{winicour1} the Bondi mass $M_B(u)$ is given by

\begin{widetext}
{\small
\begin{eqnarray}
& & M_B(u) = \int_{-1}^1\,\omega^{-1}\,\Big\{\frac{1}{2}M + \frac{1}{4}\mathrm{e}^{-2K}[(1-x^2)c^{\prime\prime}-4xc^\prime- 2c] - \mathrm{e}^{-2K}c^\prime\,(H^\prime+K^\prime)(1-x^2) - \mathrm{e}^{-2K}c\,(H^{\prime2}-2 H^\prime K^\prime-K^{\prime2})\times \nonumber \\
& & - \mathrm{e}^{-2K}c\,(H^{\prime2}-2 H^\prime K^\prime-K^{\prime2})(1-x^2) - \frac{1}{2}\mathrm{e}^{-2K}c\,[(1-x^2)(H^{\prime\prime}-K^{\prime\prime})-4x(H^\prime+K^\prime)]\Big\}dx.\nonumber\\
\label{eq34}
\end{eqnarray}}
\end{widetext}

\begin{figure}[htb]
\rotatebox{0}{\includegraphics*[scale=0.28]{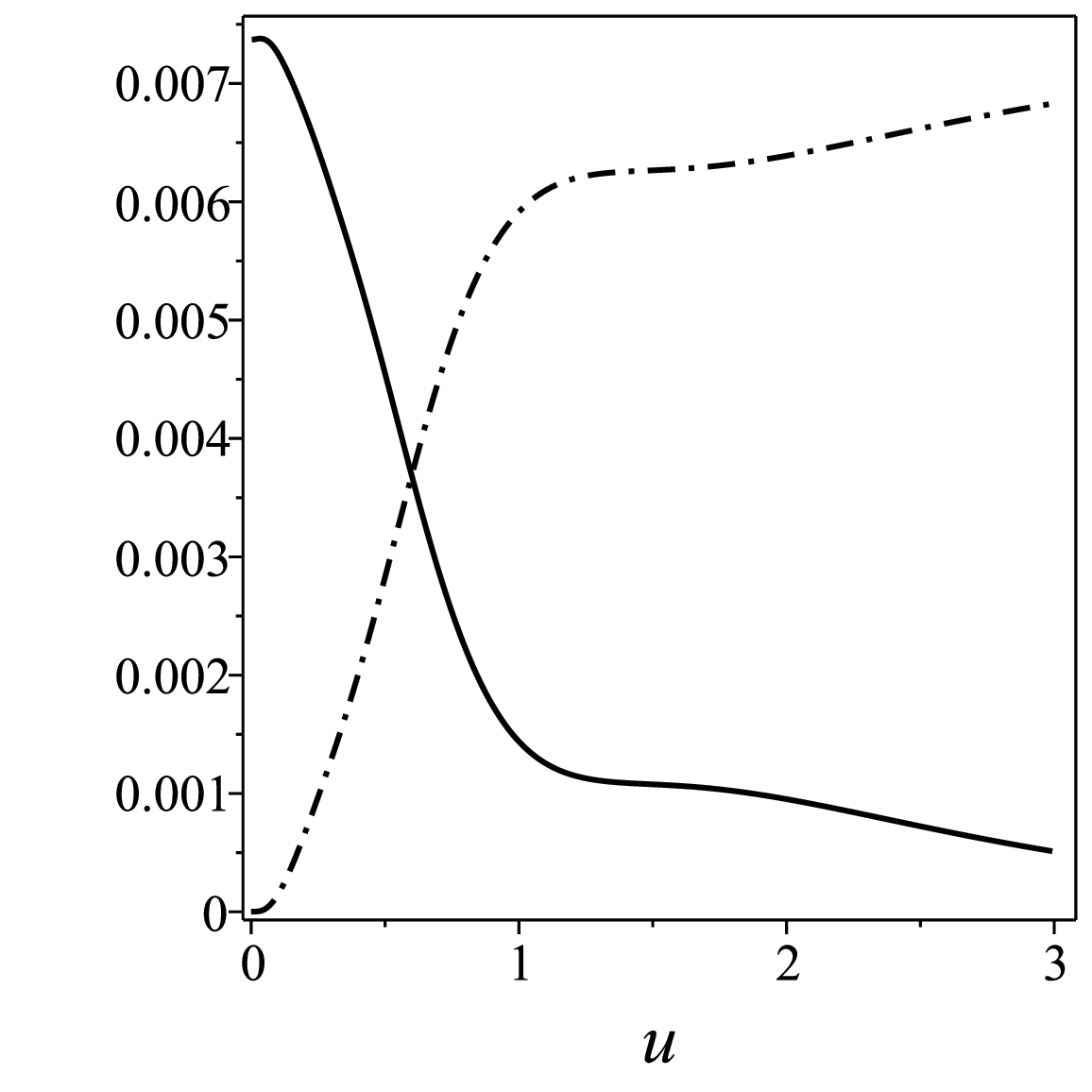}}
\rotatebox{0}{\includegraphics*[scale=0.28]{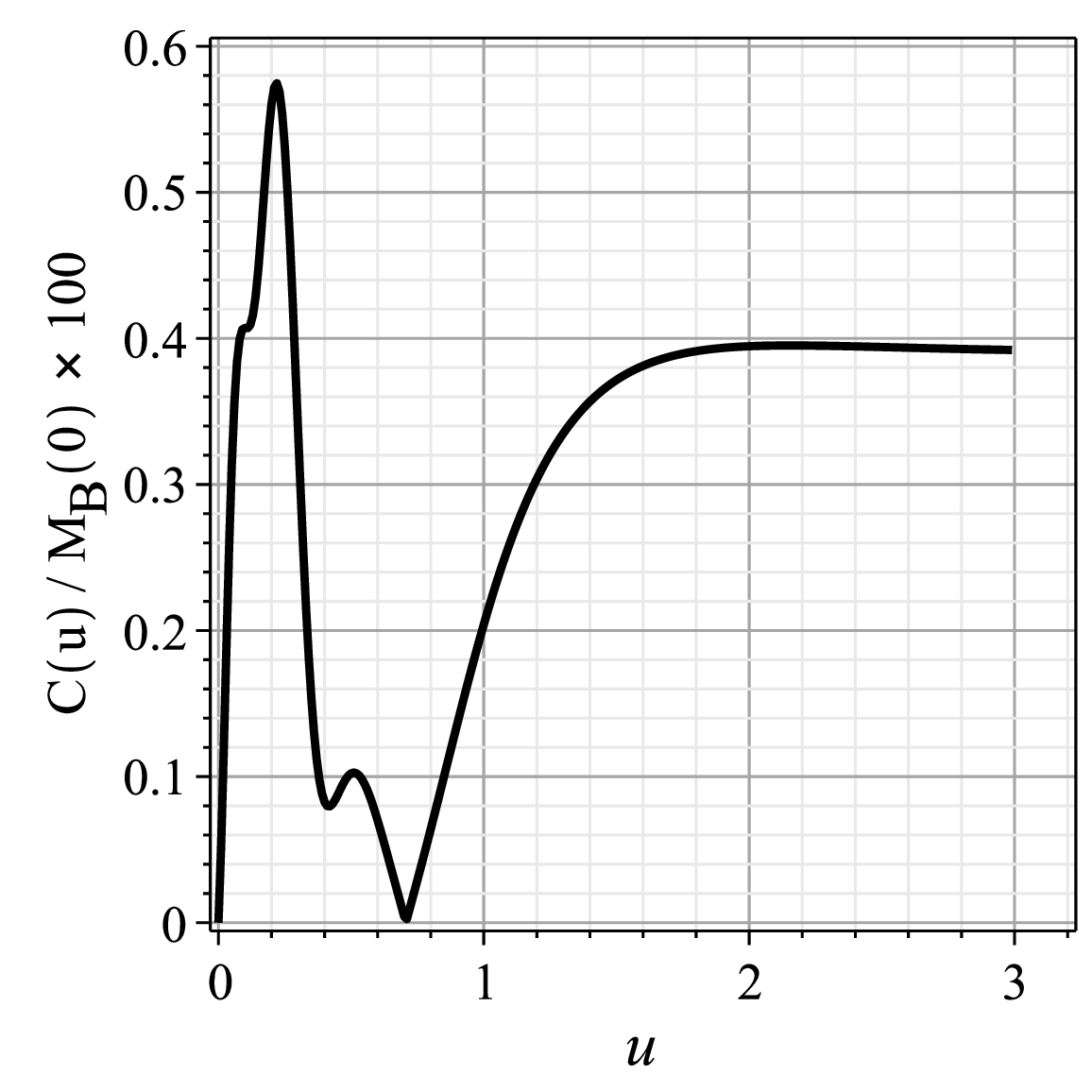}}
\caption{(a) Decay of the Bondi mass (continuous line) for the initial data (\ref{eq35}) where $A_0=0.1$ and the integral of the news function (dashed-dotted line) $1/2\,\int_{u_0}^u du\,\int_{-1}^1 \frac{\mathrm{e}^{2H}}{\omega} N^2 dx$ that measures the mass radiated by gravitational waves. (b) The relative error $|C(u)|/M_B(0) \times 100$ in the global energy conservation is evaluated up to $u=3.0$ for which more that $99.0\%$ of the initial mass has been radiated. The energy conservation is attained to about $0.58\%$ accuracy for $N_r=16,N_x=8,M_x=N_x+1$ and $M_r=2M_x$.}
\end{figure}

\noindent The standard Bondi frame~\cite{bondi1} is characterized by a choice of a coordinate system for which $H=K=L=0$ and consequently $\omega=1$. Also, in this frame the Bondi mass depends only on $M(u,x)$ and Eq. (\ref{eq29}) agrees with the original expression for the Bondi formula \cite{bondi1}. It must be remarked that all quantities listed above can be read off directly from the asymptotic expansions of the approximate expressions Eqs. (\ref{eq12}), (\ref{eq14}-\ref{eq16}).

We have checked the global energy conservation using several initial data with compact support, as for instance,

\begin{eqnarray}
&\bar{\gamma}_0(r,x) = 216 A_0\frac{r^6 (1-x^2)^2}{(1+2 r)^9},\label{eq35}\\
\nonumber \\
&\bar{\gamma}_0(r,x) = A_0 \frac{r^2 \mathrm{e}^{-(r-0.1)^2} (1-x^2)^2}{(1+r)^2} , \label{eq36}&
\end{eqnarray}


\noindent where $A_0$ is a free parameter. In Fig. 4(a) we show the decay of the Bondi mass (continuous line), $M_B(u)$, together with increase of the integral of the news function (dashed-dotted line), $1/2\,\int_{u_0}^u du\,\int_{-1}^1 \frac{\mathrm{e}^{2H}}{\omega} N^2 dx$. In fact this behavior is expected since the integral of the news function represents the amount of mass that is radiated away. We have considered the initial data (\ref{eq35}) in which $A_0=10$, $N_r=16,N_x=8$ and $M_x=N_x+1,M_r=2 M_x$. The relative error in energy conservation at each instant given by $|C(u)|/M_B(0)\times 100$, is presented in Fig. 4(b). As indicated, the energy conservation is attained to about $0.58\%$ accuracy.

\begin{figure}[htb]
\rotatebox{0}{\includegraphics*[scale=0.28]{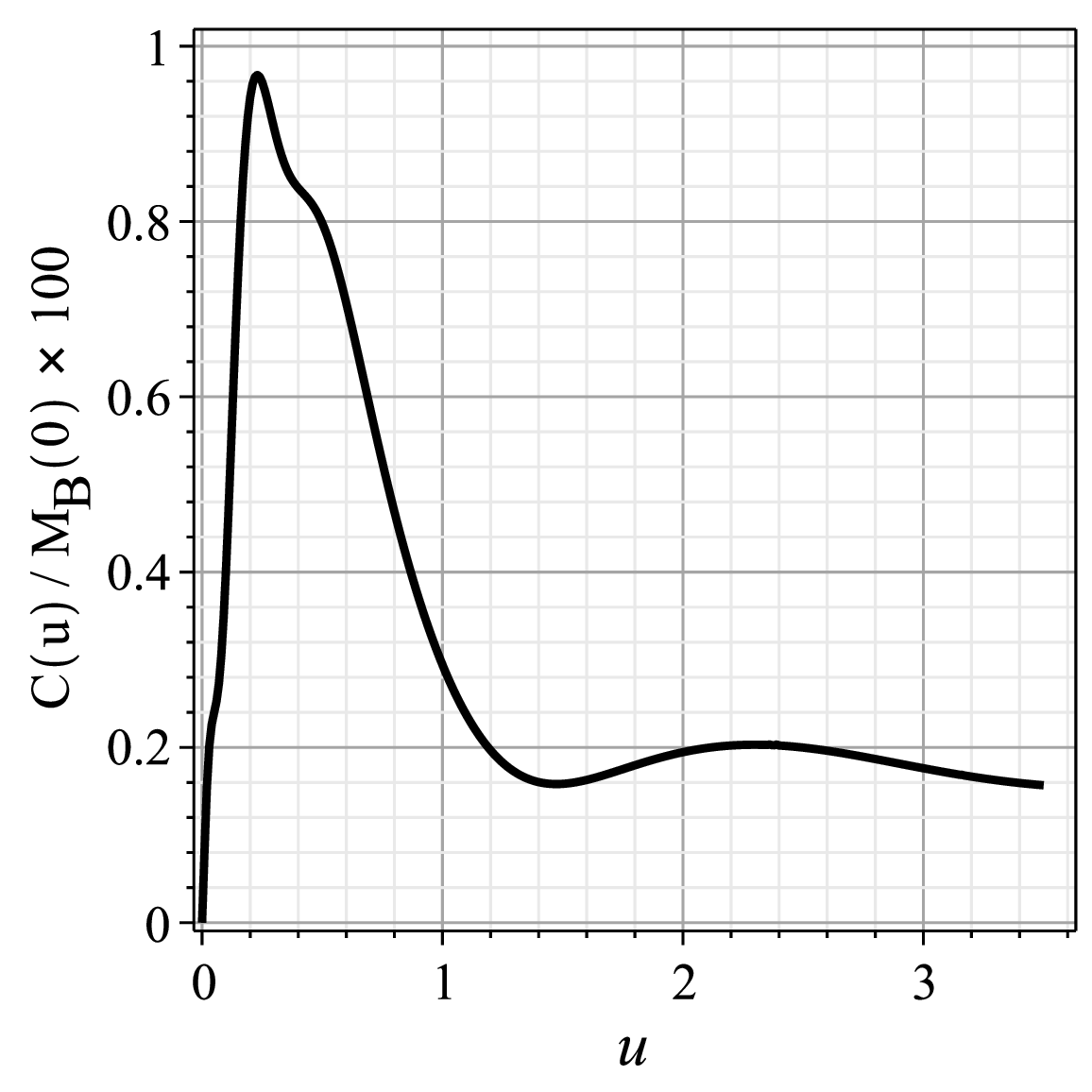}}
\rotatebox{0}{\includegraphics*[scale=0.28]{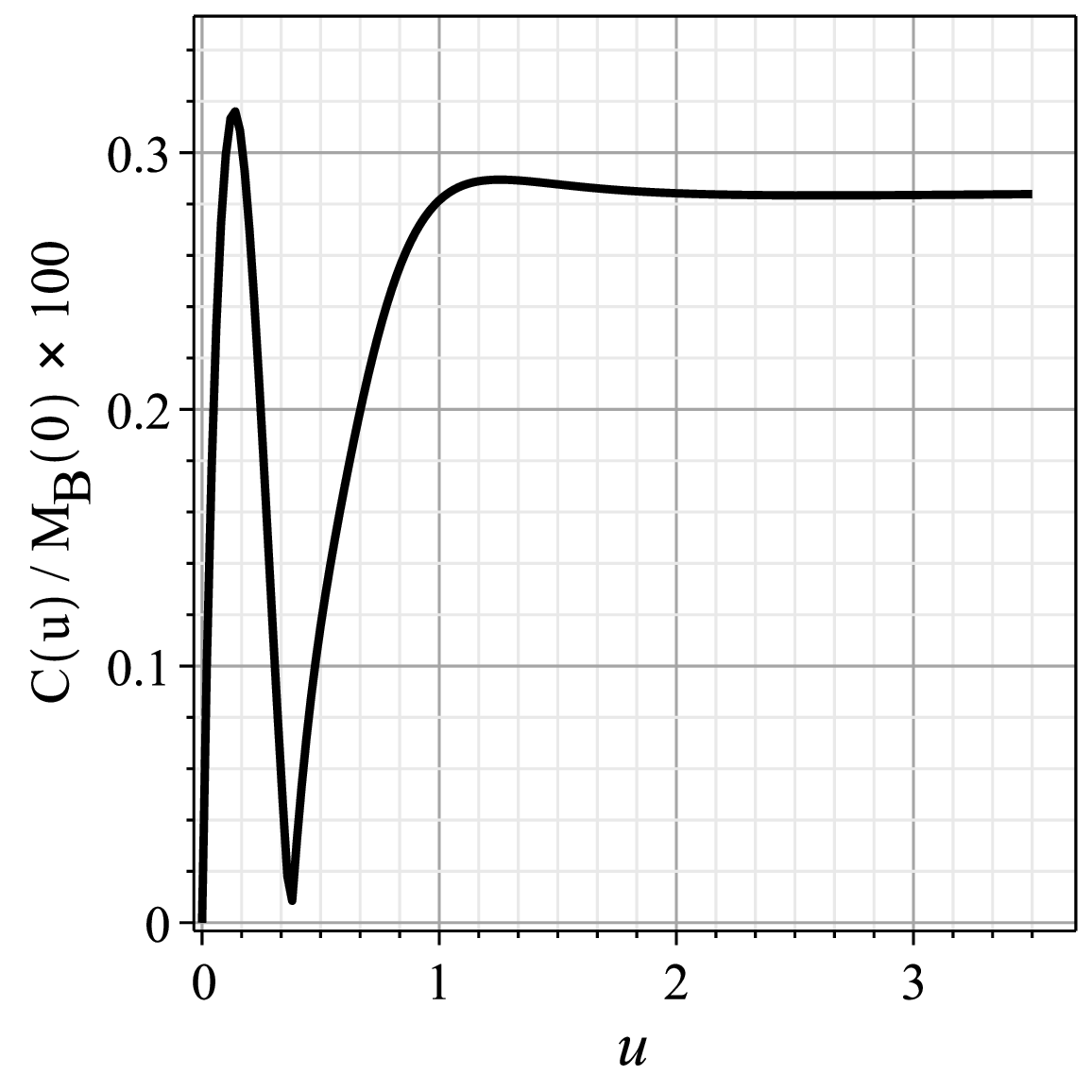}}
\caption{Plots of the relative error $|C(u)|/M_B(0) \times 100$ for the (a) Gaussian initial data (\ref{eq35}) with $A_0=0.1$, and for the initial data $\gamma_0(r,x) = 36 A_0(0.9-x)^2(1-x^2)r^4/(1+r)^9$ with $A_0=1$ that display a richer angular structure. In both cases the relative error in energy conservation is attained bellow to $1.0\%$ accuracy.}
\end{figure}

Two more plots showing the accuracy of the global energy conservation are shown in Fig. 5. In Fig. 5(a) we have considered the Gaussian initial data (\ref{eq36}) with $A_0=0.1$ and the following  truncation orders: $N_r=29,N_x=6$, $M_x=N_x+2,M_r=N_r+2$. For the second plot the following initial data with a richer angular structure as considered,

\begin{equation}
\gamma_0(r,x) = 36 A_0\frac{(0.9-x)^2(1-x^2)r^4}{(1+r)^9}
\end{equation}

\noindent with $A_0=1$, and we have used the same truncation orders indicated in Fig. 4. Notice that in both cases the maximum error in energy conservation is below $1.0\%$.

We present in Fig. 6 a sequence of 3D plots of $\bar{\gamma}(u,x,y)$ evaluated at several times starting with the Gaussian initial data (\ref{eq36}) with $A_0=0.1$. The initial data evolve until is completely radiated away after some oscillatory behavior with a rich angular structure.

\section{Discussion}

In this paper we have constructed a numerical code based on the Galerkin and Collocation method to evolve axisymmetric vacuum spacetimes described by the Bondi problem. The first task was to establish the Galerkin expansions for the metric functions $\beta,\gamma,U$ and $V$. We selected sets of basis functions such that each component satisfies the appropriate boundary conditions in accordance with the Galerkin method. The Legendre polynomials are the common angular basis functions for all expansions (see Eqs. (\ref{eq12}), (\ref{eq14}-\ref{eq16})), whereas different choices for the radial basis functions were done since the boundary conditions are not the same (see Appendix A).

The field equations are constituted by three hypersurface equations and one evolution equation. We have applied the Collocation method for the hypersurface equations to reduce them to sets of algebraic equations for the grid values  $(\bar{\beta}_{,y})_{kj},(\bar{U}_{,y})_{kj}$ and $S_{kj}$ for which the approximate metric functions $\beta_a,U_a$ and $S_a$ are reconstructed at each time. The grid points are schematically indicated in Fig. 1. This procedure allows to use higher truncation orders necessary to achieve the desirable accuracy. For the evolution equation we have applied the G-NI method taking advantage the grid values of the metric functions and their derivatives to evaluate the integrals (\ref{eq18}) via quadrature formulas. Then, the algorithm is constituted by three sets of algebraic equations and one set of ordinary differential equations for the modes $a_{kj}(u)$. We have integrated the dynamical system with a second order Runge-Kutta integrator.

The numerical tests confirmed the rapid convergence of the code with the exponential decay the error as shown in Figs. 2 and 3. The standard test of reproducing the linearized exact solution of the field equations showed that $M_x \geq N_x+1$ is the only mandatory restriction between the angular truncation orders to guarantee linear stability. This relation is a consequence of the use of a common angular basis function in the expansions of $\bar{\gamma}$ and $\bar{U}$ that can be obtained analytically from the linear version of the hypersurface equation (\ref{eq3}). The other tests explores the nonlinear domain, more specifically, we have considered the interior SIMPLE solution and the global energy conservation with several initial data. In all cases we have considered a moderated number of collocation points and modes, but they can be increased according with the necessity of describing more accurately the evolution of the spacetime.

We point out some possible directions of the present work. The main idea is to use the present numerical scheme in the study of axisymmetric collapse of matter, for instance, a scalar field where the issue of critical phenomena is of interest. Another possibility is to develop a spectral code using the G-NI and collocation methods to integrate the field equations of the Bondi-Sachs problem~\cite{sachs}, which is a 3D problem and therefore suitable to describe  general gravitational wave emission from a bounded source.

\begin{widetext}
\begin{center}
\begin{figure}[htb]
\rotatebox{0}{\includegraphics*[scale=0.7]{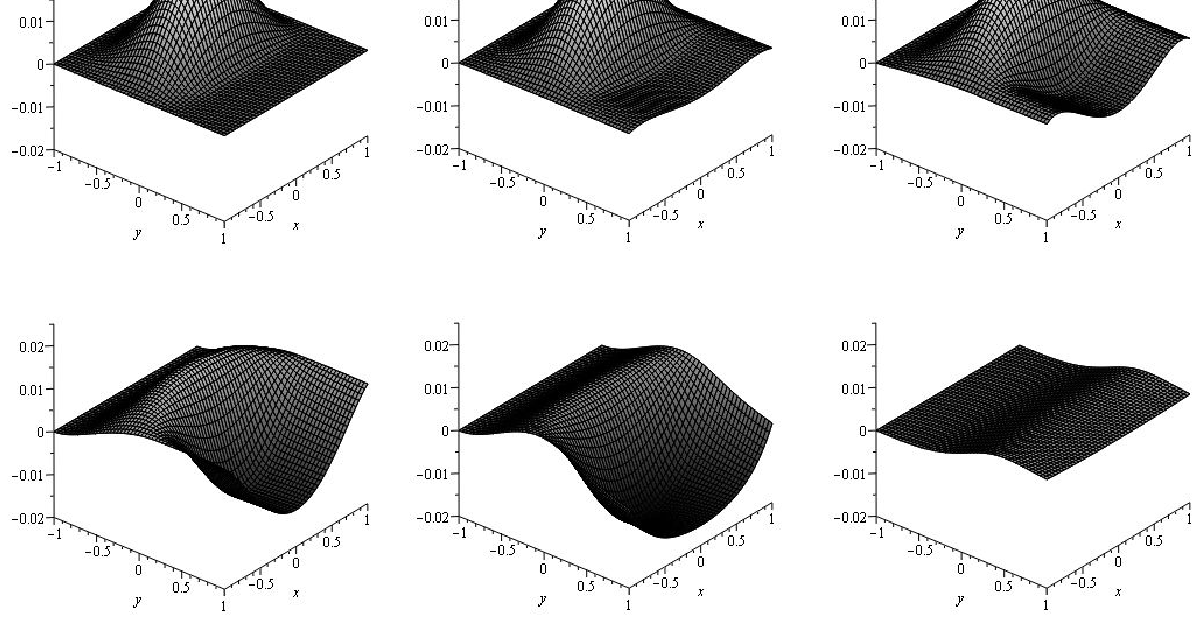}}
\vspace{-12cm}
\caption{Sequence of the numerical evolution of $\bar{\gamma}(u,x,y)$ evaluated at several times. The initial data is given by (\ref{eq36}) and the truncation orders are $N_r=29,N_x=6,M_x=N_x+2$, and $M_r=N_r+2$.}
\end{figure}
\end{center}
\end{widetext}

\appendix

\section{Radial basis functions}

We present here the radial basis functions $\Psi_k^{(\gamma)}$, $\Psi_k^{(U)}$ and $\Psi_k^{(\beta)}$. These functions are constructed using linear combinations of the rational Chebyshev polynomials \cite{boyd} $TL_k(r)$ defined in the semi-infinite range $0 \leq r < \infty$ that satisfy the boundary conditions (\ref{eq6}), (\ref{eq8}-\ref{eq11}). The radial basis functions for the Galerkin expansions of the functions $\gamma$, $U$ and $\beta$ are given by

\begin{eqnarray}
& & \Psi_k^{(U)}(r) = \frac{1}{2}\psi_k(r) \\
\nonumber \\
& &\Psi_k^{(\gamma)}(r) = \frac{1}{4}\left(\frac{1+2k}{3+2k}\psi_{k+1}(r)+\psi_k(r)\right),\\
\nonumber
& &\Psi_k^{(\beta)}(r) = \frac{1}{4}\psi_k(r) + A_k\psi_{k+1}(r) +B_k \psi_{k+2}(r)+\\
\nonumber \\
& &C_k\psi_{k+3}(r)+D_k\psi_{k+4}(r)
\end{eqnarray}

\noindent where $\psi_k(r) = TL_{k+1}(r)+TL_{k}(r)$, and

\begin{eqnarray}
A_k &=&\frac{(2k^3+14k^2+25k+9)}{4(7+6k+k^2)(k+3)}\\
\nonumber \\
B_k &=&-\frac {3(k+1)}{4(k^2+6k+7)(k+3)(2k+7)} \\
\nonumber \\
C_k &=&-\frac{(k+1)(4k^4+38k^3+118k^2+137k+48)}{4(k^2+6k+7)(k+3)(k+4)(2k+7)} \\
\nonumber \\
D_k&=&-\frac{(k+1)(3+2k)(2+4k+k^2)(k+2)}{4(7+6k+k^2)(k+3)(k+4)(2k+7)}.
\end{eqnarray}

\section{Linearized wave solutions}

The flat scalar wave equation $\Box \Phi = 0$ has the following regular solution at the hypersurface $u=0$:

\begin{equation}
\Phi_l(u,r,x) = \frac{r^l P_l(x)}{(u+1)^l (u+2r+1)^{l+1}},
\end{equation}

\noindent where $n \geq 2$. According to Ref. \cite{winicour} the solutions for $\gamma$ and $U$ are determined from two quantities, $\alpha(u,r,x)$ and $Z(u,r,x)$, by (see also \cite{newman_penrose}),

\begin{eqnarray}
\gamma = (1-x^2) \alpha_{,xx},\;\ U = -\sqrt{1-x^2} Z_{,x},
\end{eqnarray}

\noindent where,

\begin{eqnarray}
r^2 \alpha_{,r} &=& (r^2 \Phi)_{,r},\\
\nonumber \\
r^2 Z_{,r} &=& -2 [(1-x^2)\Phi_{,x}]_{,x} +4\Phi.
\end{eqnarray}

After substituting the solution (B1) into Eqs. (B3) and (B4), the solutions for  $\gamma$ and $U$ with harmonics $l=2,3...$ are obtained straightforwardly from Eqs. (B2) as:

\begin{eqnarray}
\gamma_{[l]}(u,r,x) &=& \frac{(1-x^2)[(l+2)(u+1)+4r] r^l P^{\prime\prime}_l(x)}{l (u+1)^{l+2}(u+2r+1)^{l+1}} \\ \nonumber\\
U_{[l]}(u,r,x) &=& -\frac{2\sqrt{1-x^2}[l(u+1)+2r] r^{l-1}(l+2)P^{\prime}_l(x)}{ (u+1)^{l+3}(u+2r+1)^l},  \nonumber \\
\\ \nonumber
\end{eqnarray}

\noindent where prime means derivative with respect to $x$.

\begin{acknowledgments}
The authors acknowledge the financial support of the CNPq, CAPES and FAPERJ.
\end{acknowledgments}

\end{document}